# Study on stability of the first kind of soft-matter quasicrystals

Tian-You Fan[1*] and Zhi-Yi Tang[2]
1 School of Physics, Beijing Institute of Technology, Beijing 100081,China
2 School of Computer Science and Technology, Beijing Institute of Technology, Beijing 100081,China

*Corresponding author, e-mail:tyfan2013@163.com

**Abstract** Based on extended free energy of soft-matter quasicrystals and the variation principle on thermodynamic stability, this study reports the results on stability of the first kind of soft-matter quasicrystals. They are dependent only upon the material constants, and quite simple and intuitive, the material constants can be measured by experiments. The results are significant in studying thermodynamics of the matter.

## 1 Introduction

Different from the solid quasicrystals the formation mechanism of soft-matter quasicrystals is self-assembly from spherical building blocks by supramolecules, compounds and block copolymers and so on, which is associated with chemical process [1,2]. Contrary to the solid quasicrystals, the source of stability of soft-matter quasicrystals remains a question of great debate to this day [3]. Lifshitz and his group are the pioneer on study of the stability of soft-matter quasicrystals. Following Lifshitz et al there are other groups carry out the work. The basis of their study is effective free energy proposed by Lifshitz and Petrich [4]. Different from approach developed by Lifshitz, his co-workers and followers Fan and Tang suggested extended free energy and a principle of variation of soft-matter quasicrystals [5-7] and directly find the quantitative and analytic solutions on the stability of 12- and 18-fold symmetry soft-matter quasicrystals, respectively. It is well-known, the 12- and 18-fold symmetry soft-matter quasicrystals belong to different kinds of quasicrystals, i.e., the former to the first kind of two-dimensional quasicrystals, and the latter to the second ones, respectively. Their stabilities present different characters to each other. Even if in the same kind of quasicrystals, for example, the 12-fold symmetry and 10-fold symmetry quasicrystals will present different behaviour between their stability to each other. In this paper we try to give a uniformly and systematically study to the analytic solutions and explore the stability behavior of the first kind of soft-matter quasicrystals, the results of the stability criteria are quite simple and intuitive and present bright physical meaning.

## 2 The first kind of soft-matter quasicrystals

In Refs [1,2] the first kind of soft-matter quasicrystals which include observed 12-fold and possible 5-, 8- and 10-fold symmetry ones are systematically studied from angles of symmetry, symmetry braking and elementary excitation. This kind of soft-matter quasicrystals shares the same symmetry with those of solid 12-, 5-, 8- and 10-fold symmetry quasicrystals, but due to the complex fluid naturethey have another elementary excitation i.e., the fluid phonon, its field is fluid velocity field **V** ,apart from having the phonons (the corresponding field is denoted by **u** ), and the phasons (the associated field is denoted by **w** ), respectively.  These two-dimensional quasicrystals have the field variables $u_x, u_y, u_z$; $w_x, w_y, w_z = 0$; $V_x, V_y, V_z$ if we take $z$-axis as the $n$-fold axis ($n=5,8,10,12$) in three-dimensional case.

To describe distribution, deformation and motion of soft-matter quasicrystals we need the following geometry quantities or kinetic quantities[1,2]

$$\varepsilon_{ij} = \frac{1}{2}\left(\frac{\partial u_i}{\partial x_j} + \frac{\partial u_j}{\partial x_i}\right), w_{ij} = \frac{\partial w_i}{\partial x_j}, \dot{\xi}_{ij} = \frac{1}{2}\left(\frac{\partial V_i}{\partial x_j} + \frac{\partial V_j}{\partial x_i}\right) \quad (1)$$

in (1) the first is called phonon strain tensor, the second phason strain tensor and the third the deformation rate tensor of fluid phonon, respectively.

We have the following constitutive relations[1,2]

$$\left.\begin{array}{l} \sigma_{ij} = C_{ijkl}\varepsilon_{ik} + R_{ijkl}w_{kl}, \\ H_{ij} = K_{ijkl}w_{ij} + R_{klij}\varepsilon_{kl}, \\ p_{ij} = -p\delta_{ij} + \sigma'_{ij} = -p\delta_{ij} + \eta_{ijkl}\dot{\xi}_{kl}, \end{array}\right\} \quad (2)$$

in which $\sigma_{ij}$ called the phonon stress tensor, $C_{ijkl}$ the phonon elastic tensor, and $H_{ij}$ the phason stress tensor, $K_{ijkl}$ the phason elastic tensor, $R_{ijkl}$ and $R_{klij}$ the phonon-phason coupling elastic tensor, $p_{ij}$ the fluid stress tensor, $p$ the fluid pressure, $\sigma'_{ij}$ the viscosity fluid stress tensor, $\eta_{ijkl}$ the viscosity coefficient tensor of fluid, respectively. All of independent and nonzero elements of these elastic constants $C_{ijkl}$, $K_{ijkl}$, $R_{ijkl}$ and $R_{klij}$ are determined by group representation theory, refer to Hu et el [7]. According to the group representation theory, for 12-fold symmetry quasicrystals, phonon-phason coupling elastic tensor $R_{ijkl} = 0, R_{klij} = 0$ due to decoupling between phonons and phasons for this type of quasicrystals, but for other type quasicrystals in two-dimensional quasicrystals, $R_{ijkl} \neq 0, R_{klij} \neq 0$ due to the phonon-phason coupling.

With the above information, we obtain the energy of the individual quasicrystal systems so the Hamiltonians of the relevant systems, which help us to derive the dynamics equations [1,2].

## 3 Extended free energy and determination of stability of soft-matter quasicrystals

The soft-matter quasicrystalsare a complex viscous and compressible fluid, there is the following extended inner energy density [2,5,6]

$$U_{ex} = \frac{1}{2}A\left(\frac{\delta\rho}{\rho_0}\right)^2 + B\left(\frac{\delta\rho}{\rho_0}\right)\nabla\cdot\mathbf{u} + C\left(\frac{\delta\rho}{\rho_0}\right)\nabla\cdot\mathbf{w} + U_{el} \quad (3)$$

where the first term denotes energy density due to mass density variation, the quantity $\frac{\delta\rho}{\rho_0}$ describes the variation of the matter density, in which $\delta\rho = \rho - \rho_0$ and $\rho_0$ the initial mass density. According to our computation in the cases of transient response and flow past obstacle of soft-matter quasicrystals [1], $\frac{\delta\rho}{\rho_0} = 10^{-4} \sim 10^{-3}$ for soft-matter quasicrystals, which describes this fluid effect of the matter and is greater

10 order of magnitude than that of solid quasicrystals (in this sense we can consider for the solid quasicrystals the effect due to $\frac{\delta\rho}{\rho_0}$ is very weak); the second term in (3) is one by mass density variation coupling phonons; the third represents that of mass density variation coupling phasons, and $A$, $B$ and $C$ the corresponding material constants, respectively. According to Ref [8] $C$ should be zero. These concepts are suggested by Lubensky et at [8] for the first time on the hydrodynamics of solid quasicrystals, and the present generalized dynamics of soft-matter quasicrystals is a heritage and development of the theory of Lubensky et al, in particular the Hamiltonian (or the energy functional)

$$\left.\begin{aligned} H &= H[\Psi(\mathbf{r},t)] \\ &= \int \frac{\mathbf{g}^2}{2\rho} d^d\mathbf{r} + \int \left[ \frac{1}{2} A \left(\frac{\delta\rho}{\rho_0}\right)^2 + B \left(\frac{\delta\rho}{\rho_0}\right) \nabla\cdot\mathbf{u} \right] d^d\mathbf{r} + F_{el} \\ &= H_{kin} + H_{\rho} + F_{el} \\ F_{el} &= F_u + F_w + F_{uw}, \qquad \mathbf{g} = \rho \mathbf{V} \end{aligned}\right\} \quad (4)$$

is drawn from Lubensky et al [8], in which $H_{kin}$ denotes the kinetic energy, $H_{\rho}$ the energy due to the variation of mass density, $F_{el}$ the elastic deformation energy consisting of contributed from phonons, phasons and phonon-phason coupling, respectively, the detailed definition will be given by (4)-(6) in the following. However, the dynamics equations of soft-matter quasicrystals are different from those of hydrodynamics of solid quasicrystals. Although there is such a case, the contribution of Lubensky et al is very important for the development of study of soft matter and soft-matter quasicrystals.

In equation (4)

$$U_{el} = U_u + U_w + U_{uw} \quad (5)$$

represents the elastic free energy density come from phonons, phasons and phonon-phason coupling such as

$$\left.\begin{aligned} U_u &= \frac{1}{2} C_{ijkl} \varepsilon_{ij} \varepsilon_{kl} \\ U_w &= \frac{1}{2} K_{ijkl} w_{ij} w_{kl} \\ U_{uw} &= R_{ijkl} \varepsilon_{ij} w_{kl} + R_{klij} w_{ij} \varepsilon_{kl} \end{aligned}\right\} \quad (6)$$

according to the constitutive law (2) of soft-matter quasicrystals where the meaning on $C_{ijkl}$, $K_{ijkl}$ and $R_{ijkl}, R_{klij}$ are given previously.

In accordance with the thermodynamics, the extended free energy density is defined by

$$F_{ex} = U_{ex} - TS \quad (7)$$

where $U_{ex}$ the extended inner energy density defined by (3), $T$ the absolute temperature and $S$ the entropy, respectively.

We suggest a lemma such as

***Lemma***

From (7) and (2),(3),(5) and (6), we have
$$S = -\frac{\partial F_{ex}}{\partial T}, \sigma_{ij} = \frac{\partial F_{ex}}{\partial \varepsilon_{ij}}, H_{ij} = \frac{\partial F_{ex}}{\partial w_{ij}}, \delta^2 F_{ex} \geq 0 \quad (8)$$
for a stable thermodynamic system, in which the second and third terms are equivalent to the elastic constitutive law of the material for degrees of freedom of phonons and phasons, and the last one is the stability condition of the matter, respectively. It is obvious that the extended inner energy density is a quadratic form of the relative variation of mass density $\frac{\delta\rho}{\rho_0}$, the strains $\varepsilon_{ij}$ and $w_{ij}$, and the coefficients $A$, $B$, $C_{ijkl}$, $K_{ijkl}$, $R_{ijkl}$ and $R_{klij}$ of the quadratic form a matrix $M$ (the detail depends on the concrete constitutive law for individual quasicrystal system, and refer to the subsequent sections), which is named as rigidity matrix similar to the classical elasticity, conventional crystallography and solid quasicrystallography, the thermodynamic stable condition can also be expressed by the positive definite condition of the rigidity matrix.

Accordingly under the condition (8), the validity of variation
$$\delta^2 F_{ex} \geq 0 \quad (9)$$
is equivalent to the positive definite nature of rigidity matrix $M$ which is formed by the coefficients of the quadratic form of extended inner energy density defined by (3).

## 4 The stability of the soft-matter quasicrystals with 12-fold symmetry

Among the first kind of soft-matter quasicrystals the 12-fold symmetry quasicrystals are observed already, whose stability criterion has been determined refer to Refs [2, 5,6], in which the constitutive law is a key and listed below [1]

$$\left.\begin{aligned}
\sigma_{xx} &= C_{11}\varepsilon_{xx} + C_{12}\varepsilon_{yy} + C_{13}\varepsilon_{zz} \\
\sigma_{yy} &= C_{12}\varepsilon_{xx} + C_{11}\varepsilon_{yy} + C_{13}\varepsilon_{zz} \\
\sigma_{zz} &= C_{13}\varepsilon_{xx} + C_{13}\varepsilon_{yy} + C_{33}\varepsilon_{zz} \\
\sigma_{yz} &= \sigma_{zy} = 2C_{44}\varepsilon_{yz} \\
\sigma_{zx} &= \sigma_{xz} = 2C_{44}\varepsilon_{zx} \\
\sigma_{xy} &= \sigma_{yx} = 2C_{66}\varepsilon_{xy}
\end{aligned}\right\} \quad (10a)$$

$$\left.\begin{aligned}
H_{xx} &= K_1 w_{xx} + K_2 w_{yy} \\
H_{yy} &= K_2 w_{xx} + K_1 w_{yy} \\
H_{yz} &= K_4 w_{yz} \\
H_{xy} &= (K_1 + K_2 + K_3) w_{xy} + K_3 w_{yx} \\
H_{xz} &= K_4 w_{xz} \\
H_{yx} &= K_3 w_{xy} + (K_1 + K_2 + K_3) w_{yx}
\end{aligned}\right\} \quad (10b)$$

$$\left. \begin{array}{l} p_{xx} = -p + 2\eta \dot{\xi}_{xx} - \dfrac{2}{3}\eta \dot{\xi}_{kk} \\[4pt] p_{yy} = -p + 2\eta \dot{\xi}_{yy} - \dfrac{2}{3}\eta \dot{\xi}_{kk} \\[4pt] p_{zz} = -p + 2\eta \dot{\xi}_{zz} - \dfrac{2}{3}\eta \dot{\xi}_{kk} \\[4pt] p_{yz} = 2\eta \dot{\xi}_{yz},\ p_{zx} = 2\eta \dot{\xi}_{zx},\ p_{xy} = 2\eta \dot{\xi}_{xy} \\[4pt] \dot{\xi}_{kk} = \dot{\xi}_{xx} + \dot{\xi}_{yy} + \dot{\xi}_{zz} \end{array} \right\} \quad (10c)$$

in which the material constants of elasticity and fluid are simplified, e.g.

$$\left. \begin{array}{l} C_{1111} = C_{11},\ C_{1122} = C_{12},\ C_{3333} = C_{33}, \\ C_{1133} = C_{13},\ C_{2323} = C_{44},\ C_{1212} = C_{66}, \\ (C_{11} - C_{12})/2 = C_{66}, \end{array} \right\} \quad (11)$$

$$\left. \begin{array}{l} K_{1111} = K_{2222} = K_{2121} = K_{1212} = K_1, \\ K_{1122} = K_{2211} = -K_{2112} = -K_{1221} = K_2, \\ K_{1122} = K_{1221} = K_{2112} = K_3, \\ K_{2323} = K_{1313} = K_4 \end{array} \right\} \quad (12)$$

and the phonon-phason coupling constants $R_{ijkl} = R_{klij} = 0$ due to decoupling between phonons and phasons for this type of quasicrystals and results in

$$U_{uw} = R_{ijkl}\varepsilon_{ij}w_{kl} + R_{klij}w_{ij}\varepsilon_{kl} = 0.$$

According to the above definition on inner energy density (3) and constitutive law (10) there is a rigidity matrix for point group 12mm soft-matter quasicrystals.

$$M_{12} = \begin{pmatrix} A & B & B & B & 0 & 0 & 0 & 0 & 0 & 0 & 0 & 0 & 0 \\ B & C_{11} & C_{12} & C_{13} & 0 & 0 & 0 & 0 & 0 & 0 & 0 & 0 & 0 \\ B & C_{12} & C_{11} & C_{13} & 0 & 0 & 0 & 0 & 0 & 0 & 0 & 0 & 0 \\ B & C_{13} & C_{13} & C_{33} & 0 & 0 & 0 & 0 & 0 & 0 & 0 & 0 & 0 \\ 0 & 0 & 0 & 0 & C_{44} & 0 & 0 & 0 & 0 & 0 & 0 & 0 & 0 \\ 0 & 0 & 0 & 0 & 0 & C_{44} & 0 & 0 & 0 & 0 & 0 & 0 & 0 \\ 0 & 0 & 0 & 0 & 0 & 0 & \tfrac{1}{2}(C_{11}-C_{12}) & 0 & 0 & 0 & 0 & 0 & 0 \\ 0 & 0 & 0 & 0 & 0 & 0 & 0 & K_1 & K_2 & 0 & 0 & 0 & 0 \\ 0 & 0 & 0 & 0 & 0 & 0 & 0 & K_2 & K_1 & 0 & 0 & 0 & 0 \\ 0 & 0 & 0 & 0 & 0 & 0 & 0 & 0 & 0 & K_4 & 0 & 0 & 0 \\ 0 & 0 & 0 & 0 & 0 & 0 & 0 & 0 & 0 & 0 & K_1+K_2+K_3 & 0 & K_3 \\ 0 & 0 & 0 & 0 & 0 & 0 & 0 & 0 & 0 & 0 & 0 & K_4 & 0 \\ 0 & 0 & 0 & 0 & 0 & 0 & 0 & 0 & 0 & 0 & K_3 & 0 & K_1+K_2+K_3 \end{pmatrix} \quad (13)$$

We have the theorem

***Theorem 1***

The positive definite condition of the rigidity matrix (13) is equivalent to variation (9), this leads to the stability criterion of soft-matter quasicrystals of 12mm point group as follows

$$\left. \begin{array}{l} A > 0,\ C_{11} - C_{12} > 0,\ A(C_{11}C_{33} + C_{12}C_{33} - 2C_{13}^2) - B^2(C_{11} + C_{12} - 4C_{13} + 2C_{33}) > 0, \\ C_{44} > 0,\ K_4 > 0,\ K_1 - K_2 > 0,\ K_1 + K_2 > 0,\ K_1 + K_2 + 2K_3 > 0 \end{array} \right\} \quad (14)$$

The proof of the theorem is straightforward, so it is omitted.

## 5 Stability of 8-fold symmetry soft-matter quasicrystals

The constitutive law for point group 8mm soft-matter quasicrystals [1]

$$\left. \begin{array}{l} \sigma_{xx} = C_{11}\varepsilon_{xx} + C_{12}\varepsilon_{yy} + C_{13}\varepsilon_{zz} + R(w_{xx} + w_{yy}) \\ \sigma_{yy} = C_{12}\varepsilon_{xx} + C_{11}\varepsilon_{yy} + C_{13}\varepsilon_{zz} - R(w_{xx} + w_{yy}) \\ \sigma_{zz} = C_{13}\varepsilon_{xx} + C_{13}\varepsilon_{yy} + C_{33}\varepsilon_{zz} \\ \sigma_{yz} = \sigma_{zy} = 2C_{44}\varepsilon_{yz} \\ \sigma_{zx} = \sigma_{xz} = 2C_{44}\varepsilon_{zx} \\ \sigma_{xy} = \sigma_{yx} = 2C_{66}\varepsilon_{xy} - Rw_{xy} + Rw_{yx} \end{array} \right\} \quad (15a)$$

$$\left. \begin{array}{l} H_{xx} = K_1 w_{xx} + K_2 w_{yy} + R(\varepsilon_{xx} - \varepsilon_{yy}) \\ H_{yy} = K_2 w_{xx} + K_1 w_{yy} + R(\varepsilon_{xx} - \varepsilon_{yy}) \\ H_{yz} = K_4 w_{yz} \\ H_{xy} = (K_1 + K_2 + K_3)w_{xy} + K_2 w_{yz} - 2R\varepsilon_{xy} \\ H_{xz} = K_4 w_{xz} \\ H_{yx} = K_3 w_{xy} + (K_1 + K_2 + K_3)w_{yx} + 2R\varepsilon_{xy} \end{array} \right\} \quad (15b)$$

$$\left. \begin{array}{l} p_{xx} = -p + 2\eta\dot{\xi}_{xx} - \dfrac{2}{3}\eta\dot{\xi}_{kk} \\ p_{yy} = -p + 2\eta\dot{\xi}_{yy} - \dfrac{2}{3}\eta\dot{\xi}_{kk} \\ p_{zz} = -p + 2\eta\dot{\xi}_{zz} - \dfrac{2}{3}\eta\dot{\xi}_{kk} \\ p_{yz} = 2\eta\dot{\xi}_{yz}, \quad p_{zx} = 2\eta\dot{\xi}_{zx}, \quad p_{xy} = 2\eta\dot{\xi}_{xy} \\ \dot{\xi}_{kk} = \dot{\xi}_{xx} + \dot{\xi}_{yy} + \dot{\xi}_{zz} \end{array} \right\} \quad (15c)$$

in which the evident difference of this quasicrystal system from the 12-fold symmetryquasicrystals lies in the coupling between phonons and phasons. We have the rigidity matrix for the point group 8mm soft-matter quasicrystals as follows

$$M_8 = \begin{pmatrix} A & B & B & B & 0 & 0 & 0 & 0 & 0 & 0 & 0 & 0 & 0 \\ B & C_{11} & C_{12} & C_{13} & 0 & 0 & 0 & R & R & 0 & 0 & 0 & 0 \\ B & C_{12} & C_{11} & C_{13} & 0 & 0 & 0 & -R & -R & 0 & 0 & 0 & 0 \\ B & C_{13} & C_{13} & C_{33} & 0 & 0 & 0 & 0 & 0 & 0 & 0 & 0 & 0 \\ 0 & 0 & 0 & 0 & C_{44} & 0 & 0 & 0 & 0 & 0 & 0 & 0 & 0 \\ 0 & 0 & 0 & 0 & 0 & C_{44} & 0 & 0 & 0 & 0 & 0 & 0 & 0 \\ 0 & 0 & 0 & 0 & 0 & 0 & \tfrac{1}{2}(C_{11}-C_{12}) & 0 & 0 & 0 & -R & 0 & R \\ 0 & R & -R & 0 & 0 & 0 & 0 & K_1 & K_2 & 0 & 0 & 0 & 0 \\ 0 & R & -R & 0 & 0 & 0 & 0 & K_2 & K_1 & 0 & 0 & 0 & 0 \\ 0 & 0 & 0 & 0 & 0 & 0 & 0 & 0 & 0 & K_4 & 0 & 0 & 0 \\ 0 & 0 & 0 & 0 & 0 & 0 & -R & 0 & 0 & 0 & K_1+K_2+K_3 & 0 & K_3 \\ 0 & 0 & 0 & 0 & 0 & 0 & 0 & 0 & 0 & 0 & 0 & K_4 & 0 \\ 0 & 0 & 0 & 0 & 0 & 0 & R & 0 & 0 & 0 & K_3 & 0 & K_1+K_2+K_3 \end{pmatrix} \quad (16)$$

and we have the theorem

***Theorem 2***

The positive definite condition of the rigidity matrix (16) is equivalent to the variation (9) and leads to the stability criterion of soft-matter quasicrystals of point group 8mm as follows

$$\left.\begin{array}{l} A>0,\ C_{11}-C_{12}>0,\ A(C_{11}C_{33}+C_{12}C_{33}-2C_{13}^2)-B^2(C_{11}+C_{12}-4C_{13}+2C_{33})>0,\ C_{44}>0, \\ K_4>0,\ K_1-K_2>0,\ K_1+K_2>0,\ K_1+K_2+2K_3>0,\ (C_{11}-C_{12})(K_1+K_2)-4R^2>0 \end{array}\right\} \quad (17)$$

The coupling constant $R$ has been taken into account in the stability criterion (17), which is the most evident difference with that of 12-fold symmetry quasicrystals.

The proof detail of the theorem is omitted because it is in straightforward manner.

## 6 Stability of 10-fold symmetry soft-matter quasicrystals

We consider the point group 10mm soft-matter quasicrystals, whose constitutive law is [1]

$$\left.\begin{array}{l} \sigma_{xx}=C_{11}\varepsilon_{xx}+C_{12}\varepsilon_{yy}+C_{13}\varepsilon_{zz}+R(w_{xx}+w_{yy}) \\ \sigma_{yy}=C_{12}\varepsilon_{xx}+C_{11}\varepsilon_{yy}+C_{13}\varepsilon_{zz}-R(w_{xx}+w_{yy}) \\ \sigma_{zz}=C_{13}\varepsilon_{xx}+C_{13}\varepsilon_{yy}+C_{33}\varepsilon_{zz} \\ \sigma_{yz}=\sigma_{zy}=2C_{44}\varepsilon_{yz} \\ \sigma_{zx}=\sigma_{xz}=2C_{44}\varepsilon_{zx} \\ \sigma_{xy}=\sigma_{yx}=2C_{66}\varepsilon_{xy}-R(w_{xy}-w_{yx}) \end{array}\right\} \quad (18a)$$

$$\left.\begin{array}{l} H_{xx}=K_1 w_{xx}+K_2 w_{yy}+R(\varepsilon_{xx}-\varepsilon_{yy}) \\ H_{yy}=K_2 w_{xx}+K_1 w_{yy}+R(\varepsilon_{xx}-\varepsilon_{yy}) \\ H_{yz}=K_4 w_{yz} \\ H_{xy}=K_1 w_{xy}-K_2 w_{yx} \\ H_{xz}=K_4 w_{xz} \\ H_{yx}=-K_2 w_{xy}+K_1 w_{yx}+2R\varepsilon_{xy} \end{array}\right\} \quad (18b)$$

$$\left.\begin{array}{l} p_{xx}=-p+2\eta\dot{\xi}_{xx}-\dfrac{2}{3}\eta\dot{\xi}_{kk} \\ p_{yy}=-p+2\eta\dot{\xi}_{yy}-\dfrac{2}{3}\eta\dot{\xi}_{kk} \\ p_{zz}=-p+2\eta\dot{\xi}_{zz}-\dfrac{2}{3}\eta\dot{\xi}_{kk} \\ p_{yz}=2\eta\dot{\xi}_{yz},\ p_{zx}=2\eta\dot{\xi}_{zx},\ p_{xy}=2\eta\dot{\xi}_{xy} \\ \dot{\xi}_{kk}=\dot{\xi}_{xx}+\dot{\xi}_{yy}+\dot{\xi}_{zz} \end{array}\right\} \quad (18c)$$

and we obtain the rigidity matrix for the 10mm point group soft-matter quasicrystals

$$M_{10} = \begin{pmatrix} A & B & B & B & 0 & 0 & 0 & 0 & 0 & 0 & 0 & 0 & 0 \\ B & C_{11} & C_{12} & C_{13} & 0 & 0 & 0 & R & R & 0 & 0 & 0 & 0 \\ B & C_{12} & C_{11} & C_{13} & 0 & 0 & 0 & -R & -R & 0 & 0 & 0 & 0 \\ B & C_{13} & C_{13} & C_{33} & 0 & 0 & 0 & 0 & 0 & 0 & 0 & 0 & 0 \\ 0 & 0 & 0 & 0 & C_{44} & 0 & 0 & 0 & 0 & 0 & 0 & 0 & 0 \\ 0 & 0 & 0 & 0 & 0 & C_{44} & 0 & 0 & 0 & 0 & 0 & 0 & 0 \\ 0 & 0 & 0 & 0 & 0 & 0 & \frac{1}{2}(C_{11}-C_{12}) & 0 & 0 & 0 & -R & 0 & R \\ 0 & R & -R & 0 & 0 & 0 & 0 & K_1 & K_2 & 0 & 0 & 0 & 0 \\ 0 & R & -R & 0 & 0 & 0 & 0 & K_2 & K_1 & 0 & 0 & 0 & 0 \\ 0 & 0 & 0 & 0 & 0 & 0 & 0 & 0 & 0 & K_4 & 0 & 0 & 0 \\ 0 & 0 & 0 & 0 & 0 & 0 & -R & 0 & 0 & 0 & K_1 & 0 & -K_2 \\ 0 & 0 & 0 & 0 & 0 & 0 & 0 & 0 & 0 & 0 & 0 & K_4 & 0 \\ 0 & 0 & 0 & 0 & 0 & 0 & R & 0 & 0 & 0 & -K_2 & 0 & K_1 \end{pmatrix} \quad (19)$$

The corresponding theorem on the stability is

**Theorem 3**

Due to the equivalence of the positive definite condition of the matrix (19) to the variation (9) leads to the stability criterion of soft-matter quasicrystals of point group 10mm as follows

$$\left. \begin{array}{l} A > 0, \ C_{11} - C_{12} > 0, \ A(C_{11}C_{33} + C_{12}C_{33} - 2C_{13}^2) - B^2(C_{11} + C_{12} - 4C_{13} + 2C_{33}) > 0, \\ C_{44} > 0, \ K_4 > 0, \ K_1 - K_2 > 0, \ K_1 + K_2 > 0, \ (C_{11} - C_{12})(K_1 + K_2) - 4R^2 > 0 \end{array} \right\} \quad (20)$$

The proof of the theorem is omitted due to the straightforward manner.

## 7 Conclusion and discussion

By a complete different approach compared with Ref [3] this paper discussed the stability of the first kind soft-matter quasicrystals, which directly is based on the thermodynamics with the help of generalized dynamics of the matter, and offered some quantitative and very simple results, the stability depends only upon the material constants, which can be measured by experiments.

The stability of soft-matter quasicrystals is related with the fluid properties of the matter which described by material constants $A$ and $B$ in additional to the elasticity properties described by $C_{ij}$, $K_{ij}$ and $R_{ij}, R_{ji}$, the latter depends upon the symmetry structures of the quasicrystal systems, respectively. This leads to the evident differences of the stability criteria between 8-, 10-fold and 12-fold symmetry quasicrystals, because the former present phonon-phason coupling and the latter present phonon-phason decoupling. There are some differences too between 8- and 10-fold symmetry quasicrystals due to the differences of their structures.

The discussion can also be given in accordance with the first law of thermodynamics, omitting the detail, in this law we have

$$\sigma_{ij} = \frac{\partial U_{ex}}{\partial \varepsilon_{ij}}, H_{ij} = \frac{\partial U_{ex}}{\partial w_{ij}}, \delta^2 U_{ex} \geq 0 \quad (21)$$

where $U_{ex}$ is defined by (3) (of course $C=0$), which is a quadratic form, the condition $\delta^2 U_{ex} \geq 0$ requires the rigidity matrix must be positive definite, so leads to the results (14),(17) and (20) respectively, i.e., the theorems hold.

The stability of 18-fold symmetry soft-matter quasicrystals, which belong to the second kind of soft-matter quasicrystals, was discussed by Ref [6], the general results on stability of this kind of soft-matter quasicrystals will be given uniformly and systematically by another report of ours.

**Acknowledgement** The work is supported by the National Natural Science Foundation of China through the grant 11272053. The first author of the article thanks Prof R Lifshitz of Tel Aviv University for presenting the electronic copy of Ref [3]. Zhi-Yi Tang is grateful to the support in part of the National Natural Science Foundation of China through the grant11871098.